\documentclass[prl,twocolumn,aps,showpacs]{revtex4}
\usepackage{graphicx}
\begin{document}
\title{Orientation Dependence of Step Stiffness: Failure of SOS and Ising Models to
Describe Experimental Data}
\vspace{-0.3cm}
\author{Sabine Dieluweit, Harald Ibach and Margret Giesen}
\email{m.giesen@fz-juelich.de
}

\affiliation{Institut f\"ur Schichten und Grenzfl\"achen, ISG 3,
Forschungszentrum J\"ulich, D 52425 J\"ulich, Germany}
\author{T. L. Einstein}
\affiliation{Department of Physics, University of Maryland,
College Park, Maryland 20742-4111, USA}

\date{\today}
\begin{abstract}

We have investigated the step stiffness on Cu(001) surfaces as a function of step
orientation by two independent methods at several temperatures near 300 K. Both sets
of data agree well and show a substantial dependence of the stiffness on the
angle of orientation. With the exception of steps oriented
along $\langle 110\rangle$, the experimental stiffness is significantly larger than the
stiffness calculated within the solid-on-solid (SOS) model and the Ising-model, even
if next nearest-neighbor interactions are taken into account. Our results have
considerable consequences for the understanding and for the theoretical modeling of
equilibrium and growth phenomena, such as step meandering instabilities.
\end{abstract}
\maketitle


Self-assembly of nanostructures is a key route on the roadmap
of device technology at ever-shrinking lengths. Several
methods for self-assembly of nanostructures via kinetics, kinetic instabilities, or as
equilibrium phenomena are associated with the properties of steps on solid surfaces
\cite{T}. In the context of nanostructures the step stiffness $\tilde{\beta}$---which
describes the resistance of a step to meandering---is one of the key parameters in the
widely applicable step continuum model \cite{JW}.
As the free energy cost per
length (along the macroscopic edge direction) for elongating the microscopic length of a
step due to fluctuations, $\tilde{\beta}$ plays a central role in virtually all properties
involving step excitations.  For steps in equilibrium, the experimental and
theoretical work focussed on spatial and temporal fluctuations and the repulsive or
attractive interactions between fluctuating steps
\cite{JW,G}.  When steps represent a growth front, stable step-flow growth occurs due to
the presence of an Ehrlich-Schwoebel (ES) barrier
\cite{EH,SS} which hinders interlayer diffusion. This barrier, however, can also give
rise to a step meandering instability: Bales and Zangwill \cite{BZ} showed that this
form of instability selects a fastest-growing mode of meandering, which determines the
morphology of the surface in the long-time limit. More recently it was shown that
meandering can also be caused by a kink Ehrlich-Schwoebel barrier (KESE)
\cite{PDE,KKK}. Experimentally, a meandering instability was observed on Cu(001)
vicinals by Maroutian et al.\ \cite{MDE}. The existing theories, while
accounting qualitatively for the effect, fail to give such quantitative aspects
as the temperature-dependence of the dominant wave length \cite{MDE}.

Regardless of the precise cause of the instability, the forces
driving the step into meandering compete against diffusion currents
tending to smooth out gradients in the chemical potential $\mu$ along steps. The
chemical potential is the product of $\tilde{\beta}$ and the curvature $\kappa$. Thus, it
is the stiffness which keeps deviations from the mean direction low. For steps in
equilibrium, such as on vicinal surfaces in the absence of growth, the stiffness
determines the magnitude of the equilibrium fluctuations of steps and thereby the
fluctuations in the terrace-width distribution.

Unlike the surface stiffness, the step stiffness has a defined value for any
orientation since a step is thermodynamically rough above temperature $T \! = \! 0$.
Present analytical theories of the meandering instabilities merely consider an angle
independent (``isotropic") stiffness
\cite{KKK,MDE,PMS,GPM,KK,P} and estimate $\tilde{\beta}(\theta)$ as its value for a
step orientation along the direction $\theta \! =\! 0$ of close packing (cf. Eq.\
(\ref{eq:therm}) below). In Monte-Carlo simulations (e.g. Ref.\ \cite{KKK}), where the
modeling of the step meandering is based on the solid-on-solid (SOS) model \cite{K}, a
particular angular dependence of the stiffness is automatically built in by considering an
angle dependent line tension (step free energy per length)
$\beta(\theta)$ (since $\tilde{\beta}(\theta) = \beta(\theta) +
\beta^{\prime\prime}(\theta)$).

In this Letter
we show that not only the standard assumption of isotropic stiffness but even the
angular-dependent step stiffness of the SOS or Ising model \cite{WPW} is far from reality
for the much-studied \cite{JW,G} Cu(001) surface, typical of late-transition/noble metal
surfaces and on which the meandering instability was observed \cite{MDE}: Our
experimentally determined stiffness varies by several orders of magnitude for orientations
close to
$\langle 110\rangle$, in agreement with the SOS and Ising models. For larger angles,
however, the experimental stiffness is about quadruple the value calculated in the two
models \cite{RW,ZA}. As we will argue, this deviation cannot be accounted for by
including next-nearest-neighbor interactions. Rather it must be attributed to
many-body interactions such as kink-kink interactions and/or effective corner
energies.

Our step-stiffness data come from two independent experiments. One set of data is obtained
by measuring the step-step distance correlation function for steps which are misoriented by
an azimuthal angle $\theta$ from the high-symmetry, close-packed $\langle 110 \rangle$
direction and therefore contain forced kinks. The second set is obtained by analyzing
equilibrium shapes of two-dimensional islands
\cite{SGI,GSI}.  Here $\theta$ denotes the angle between the local island-edge orientation
and the $\langle 110 \rangle$ direction (so that $\tan \theta$ corresponds to the local
slope) rather than (as in Ref.\ \cite{SGI}, e.g.) the angle denoting the position on the
perimeter \cite{SVK}.

The experiments were performed in a standard ultra-high vacuum (UHV)
chamber equipped with a variable-temperature microscope
\cite{B87,FWB} with high thermal drift stability.  For the experiments on the step-step
correlation function we used copper surfaces with Miller indices  (5,8,90). Hence, the
surfaces are tilted relative to the (001) surface by the polar angle $\phi$=
5.98$^{\circ}$.  The resulting monatomic steps on the surface are rotated with respect to
the close-packed $\langle 110 \rangle$ direction by a nominal angle $\theta \! = \!
12.99^{\circ}$, which corresponds to a density 0.23 of geometric kinks. The accuracy of
orientation is naturally limited by the mosaic structure of the single crystal.
Crystal
cleaning procedures followed established recipes \cite{GSI} after which the
sample surfaces revealed clean parallel steps, which, however, merely on the
average display the nominal rotation angle of 12.99$^{\circ}$ of the (5,8,90)
surface. The misalignment $\theta$ of the steps with respect to the $\langle 110
\rangle$-direction varied locally
between 0$^{\circ}$ and 45$^{\circ}$. The size of the local area of constant
angle was of order 10$^4$ nm$^2$. Fig.1 shows an STM image of the
Cu(5,8,90) surface at 301 K with $\theta \! = \! 7.13^{\circ}$, i.e. with an average
concentration 0.13 of forced kinks. The mean value of $\theta$ of the steps was
determined by averaging over all steps in the image. As is obvious from Fig. 1,
the $\langle 110 \rangle$ direction (black solid arrow in Fig. 1) is easily
defined in an STM image by the straight step segments between individual,
monatomic kinks.

\begin{figure}
\includegraphics[width=3in]{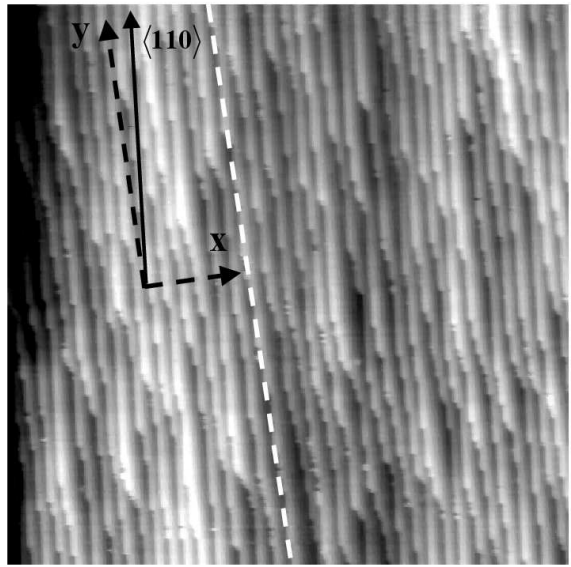}
\caption{\label{fig1}STM image of a surface region on Cu(5,8,90)
where on the average the steps run 7.13$^{\circ}$ off the
atomically dense $\langle 110 \rangle$ direction. The scan width
is 24.3 nm.}
\end{figure}

Spatial fluctuations of steps can be analyzed via a step
correlation function $G(y)$\cite{BEW}
\begin{equation}
G(y)=\langle[x(y+y^{\prime})-x(y^{\prime})]^2\rangle =
\frac{k_BT}{\tilde{\beta}a_{\parallel}}|y|a_{\parallel},
\label{eq:Gy}
\end{equation}

\noindent where $y$ and $y^{\prime}$ are coordinates parallel to the mean direction along
which the step runs (see Fig.\ 1) \cite{N}, while $a_{\parallel}$ is the atomic length
unit (along $\langle 110 \rangle$).  We have measured
the spatial correlation function $G(y)$ for various angles
$\theta$  on the  Cu(5,8,90) surface at temperatures between 293 and 328 K.
For the highest temperature and larger angles $\theta$, the plot of $G(y)$ showed a
small influence of the time dependence \cite{GJP,GI} of the fluctuations by the curving
of $G(y)$ at small $y$.  The slope in Eq.\ (1) was determined for larger $y$ where
$G(y)$ was a linear.  The resulting values for the dimensionless quantity
$k_BT/\tilde{\beta}a_{\parallel}$ are plotted in Fig.\ 2 as gray symbols.
The data points represent an average over 100 individual steps with a total step
length of 1-2 $\mu$m each. Included as stars are data at $\theta$ = 0$^{\circ}$
taken from previous work of our group  for Cu(11n) surfaces, $n$=13 and 19,
for which $k_BT/\tilde{\beta}(0)a_{\parallel}$ = 0.019 and 0.0129,
respectively \cite{GI,GSJ,GGRK}.

The step
stiffness $\tilde{\beta}$ can also be determined experimentally by analyzing
equilibrium  shapes of islands \cite{GSI}. Islands in equilibrium have a well-defined
chemical potential which is the same  all along the perimeter of an island. The
chemical potential of the island is proportional to the product of the curvature
$\kappa(\theta)$ of the island perimeter and the step stiffness
$\tilde{\beta}(\theta)$  (via the Gibbs-Thomson equation).

For islands in equilibrium, as shown previously \cite{GSI}, the chemical
potential can likewise be equated to the free energy per length $\beta_0$ of the densely
packed steps oriented along $\langle 110 \rangle$ divided by the distance of
this step segment from the center of the island $r_0$. Then one obtains

\begin{equation}
\tilde{\beta}(\theta)=\frac{\beta_0}{r_0\kappa(\theta)}.
\label{eq:bt}
\end{equation}

\noindent Since the absolute number for the step free energy
$\beta_0$ for Cu(001) is known from experiment, $\beta_0$  = 220
meV $ a_{\parallel}^{-1}$ \cite{GSI,SGV}, Eq.\ (\ref{eq:bt})  enables the
calculation of the orientation dependence of
$k_BT/\tilde{\beta}(\theta)a_{\parallel}$ from the equilibrium island shape.
Data were extracted from experiment by averaging over more than a thousand
individual island shapes which were measured at temperatures between 287
and 313 K. The temperature dependence of the equilibrium shapes is small in the
range of $\theta$ of interest here. Using the mean shapes obtained for each
temperature we have calculated the curvature as a function of the angle $\theta$
and furthermore, $k_BT/\tilde{\beta}(\theta)a_{\parallel}$  following Eq.\
(\ref{eq:bt}). The temperature-averaged results are plotted in Fig.\ 2 as solid
diamonds. They agree quite well with the data from the correlation function
$G(y)$, except for $\theta$ = 0 where it appears as if
$k_BT/\tilde{\beta}(0)a_{\parallel}$  obtained from the island equilibrium
shapes would be significantly higher. This is, however, an artifact of the island
shape analysis. Technically one cannot determine the rapidly changing curvature
of the island equilibrium shape near the $\langle 110 \rangle$ direction.

\begin{figure}
\includegraphics[width=3in]{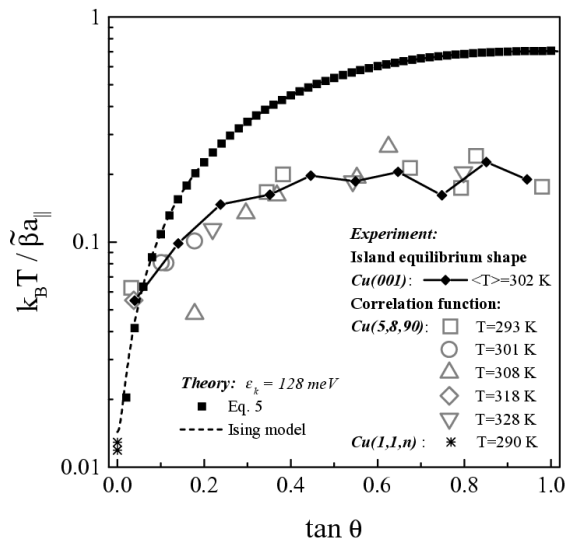}
\caption{\label{fig2}Dimensionless inverse stiffness
$k_BT/\tilde{\beta}(\theta)a_{\parallel}$  measured via spatial
step fluctuations (gray symbols) and via island equilibrium
shapes (solid diamonds). Full squares represent Eq.\
(\ref{eq:lowTstf}) and the dashed line is calculated from the
temperature-dependent analytical solution of the island
equilibrium shape in the nearest-neighbor Ising model.}
\end{figure}

The rapid descent of $k_BT/\tilde{\beta}(\theta)a_{\parallel}$ close to
$\theta$ = 0 is a real effect. It can be shown that at $\theta$ = 0 and for
moderate temperatures,
$k_BT/\tilde{\beta}(0)a_{\parallel}$ equals twice the concentration of
thermally excited kinks \cite{BEW}:

\begin{equation}
k_BT/\tilde{\beta}(0)a_{\parallel} = 2 (a_{\perp}/a_{\parallel})^2
e^{-\epsilon_k/k_BT},
\label{eq:therm}
\end{equation}

\noindent where
$\epsilon_k$ is the kink energy.  Since the unit spacing $a_{\perp}$ between $\langle 110
\rangle$ rows is the same as $a_{\parallel}$ for an (001) fcc surface, we do not
distinguish between them henceforth, calling both $a_1$.  Eq.\ (\ref{eq:therm}) is model
independent and exact if
$z \equiv \exp(-\epsilon_k/k_BT) \ll 1$.

In order to compare our experimental data for
$k_BT/\tilde{\beta}(\theta)a_1$ with theory, we make use of explicit exact
expressions for $\beta(\theta)$ for the nearest-neighbor Ising and SOS models
\cite{RW,ZA}, from which we obtain $\tilde{\beta}(\theta)$ as $\beta(\theta) +
\beta^{\prime\prime}(\theta)$.  While the general expression can be readily handled
numerically,
$T \ll \epsilon_k/k_B$ in the present problem, permitting an expansion in
$z$ (which is of order $6\! \times \! 10^{-3}$ when we specify $\epsilon_k$ below).  For
the Ising model, Rottman and Wortis
\cite{RW} found

\begin{eqnarray}
\beta(\theta) a_1/k_BT = (\epsilon/k_BT)(|\cos\theta| + |\sin\theta|) \nonumber \\-\mbox{\Large\bf
(}(|\cos\theta|+ |\sin\theta|)\ln(|\cos\theta|+ |\sin\theta|)  \nonumber \\ -
|\cos\theta|\ln(|\cos\theta|) - |\sin\theta|
\ln(|\sin\theta|)\mbox{\Large\bf )} \nonumber \\-\frac{|\cos\theta|^3 + |\sin\theta|^3}{|\cos\theta|
|\sin\theta|}z^2 +O(z^4).
\label{eq:RWIsing}
\end{eqnarray}
\noindent The energy-like first term in Eq.\ (\ref{eq:RWIsing}) makes no contribution to
$\tilde{\beta}(\theta)$ since it is canceled by its second derivative.
We then obtain the following expansion for the stiffness:

\begin{eqnarray}
k_BT/\tilde{\beta}(\theta)a_1 &=& (|\cos\theta|+|\sin\theta|)|\cos\theta||\sin\theta|
\nonumber \\+ 2(|\cos\theta|&+&|\sin\theta|)^3(|\csc\theta||\sec\theta|-1)z^2  +O(z^4).
\label{eq:zlowT}
\end{eqnarray}

\noindent The first term in Eq.\ (\ref{eq:zlowT}) increases gradually but
monotonically with $\theta$, while the monotonically-decreasing coefficient of $z^2$
is sharply peaked at $\theta \! = \! 0$.  Hence, except near $\theta \! = \! 0$
(specifically, so long as $\tan\theta \gg z$), the $z^2$ term in Eq.\ (\ref{eq:zlowT}) can
be neglected, leaving
\begin{equation}
\frac{k_BT}{\tilde{\beta}(\theta)a_1} \approx \! (\tan^2\theta +\!
|\tan\theta|)|\cos\theta|^3=
\frac{\tan^2\theta +\! |\tan\theta|}{(1+\tan^2\theta)^{\frac{3}{2}}}
\label{eq:lowTstf}
\end{equation}

\noindent The result in Eq.\ (\ref{eq:lowTstf}) can be obtained from a combinatorical
analysis for the number of ways of arranging the kinks forced by the azimuthal
misorientation $\theta$ \cite{RW,CK}. Since we expect $k_BT/\tilde{\beta}(\theta)a_1
\rightarrow 2z$ as $\theta \rightarrow 0$, the expansion in Eq.\ (\ref{eq:zlowT})
evidently fails near this limit.

The temperature independent expression for
$k_BT/\tilde{\beta}(\theta)a_1$ from Eq.\ (\ref{eq:lowTstf}) is
displayed in Fig.\ 2 as solid squares. With the help of Eq.\ (\ref{eq:bt})
$k_BT/\tilde{\beta}(\theta)a_1$ can also be calculated for arbitrary temperatures
from the analytical solution of the island equilibrium shape within the Ising model
\cite{ZA,GSI}. The result is displayed as a dashed line in Fig. 2. The Ising parameter
representing the kink energy as well as the step energy per atom was chosen as
$\epsilon_k$=128 meV \cite{GI,GSJ}.

While experimental data and theory agree well for small $\theta$, they disagree
substantially for larger $\theta$. The agreement for small $\theta$ is due to the
fact that eq. (4) is independent of any assumption other than $z \ll 1$. The
reason for the deviation at larger $\theta$ can be understood best by considering
the special case $\theta$ =45$^{\circ}$. From Eq.\ (\ref{eq:lowTstf}) one obtains

\begin{equation}
k_BT/\tilde{\beta}(45^{\circ})a_1 = 1/\sqrt{2}.
\label{eq:b45}
\end{equation}

This result can be derived stragihtforwardly by considering
the spatial correlation function $G(y)$ for a freely meandering steps in the $\langle 100
\rangle$ direction if overhangs are excluded and the statistical weight factors for
microscopic paths take only nearest neighbor broken bonds into account. As overhangs are
excluded the result is correct to terms of order $z^4$ which is an excellent approximation
for the temperatures considered here. The fact that
$k_BT/\tilde{\beta}(\theta)a_1$ is lower than predicted by Eq.\ (\ref{eq:b45}) must
therefore mean that the $\langle 100 \rangle$ oriented step ($\theta$=45$^{\circ}$;
100\% kinked) with one kink following another is energetically favored over paths which
deviate from this orientation. This is in accord with the analysis of island equilibrium
shapes \cite{GSI}, which  found that the ratio of line tensions $\beta_{\langle 100
\rangle}/\beta_{\langle 110 \rangle}$ at
$T\! =\! 0$ K is 1.24 rather than $\sqrt{2}$ as predicted in the Ising model.

We note that diagonal, next-nearest neighbor interactions characterized by
$\epsilon_{nn}$ cannot be the sole explanation.
It is straightforward to show that the experimentally observed line tensions
for the $\langle 100 \rangle$ and  $\langle 110 \rangle$-directions cannot be fit
simultaneously with the kink energy to a model with nnn-pair-potentials. As for
the temperature dependence, nnn-interactions were considered, e.g., in the
Akutsus' study of Si(001)  \cite{AA}.  To leading order, their
expansion factor
$\exp(-\epsilon_k/k_BT)$ becomes
$\exp(-(\epsilon_k+2\epsilon_{nn}/k_BT)$.  However, this change only affects the
negligible $z^2$ term; there is no change to Eq.\ (\ref{eq:lowTstf}).
Thus, the lower line tension of steps at larger
angles $\theta$ and, in particular, that of a $\langle 100 \rangle$-oriented step
($\theta$=45$^{\circ}$) and the lower corresponding spatial step fluctuations can
therefore not be described by a nnn-pair potential. Rather they must be
attributed to many body interactions. Such many body contributions might exist in
the form of (attractive) kink-kink interactions and/or (negative) corner energies
involved in the formation of kinks in a  $\langle 100 \rangle$-step. These
interactions would keep a meandering step with a mean orientation along $\langle
100 \rangle$ close to the central path which is also microscopically a
$\langle 100 \rangle$-direction and in which kinks follow immediately one after
another. Spatial fluctuations would therefore be smaller than in the Ising model.
Models which might produce such behavior are under investigation
\cite{GSE}.

We finally comment on the consequences
of the strong angle dependence of the stiffness on the step meandering
instability \cite{BZ,PDE,KKK,MDE,PMS,GPM,KK,P}. Under plausible assumptions,
the most unstable, fastest growing mode $\lambda_u \propto \tilde{\beta}^{1/2}$ in a linear
stability analysis.  Using then Eq.\ (\ref{eq:therm}) for the
calculation of
$\lambda_u$ not only overestimates the effective stiffness of the technically
always slightly misoriented steps, it also introduces an unrealistic temperature
dependence.  A very recent theoretical study \cite{E02} of the meandering instability of
Cu (001) did consider a step stiffness with angular dependence, but with
behavior so different \cite{noteE} from Eq.\ (\ref{eq:lowTstf}) to preclude useful
comparison.  Recognition of the remarkable angular dependence of the
stiffness is a crucial ingredient for attempts to account for the meandering
instabilities \cite{MDE} and other statistical properties of vicinals and of single-layer
islands.

\section*{Acknowledgment}

We acknowledge the high-accuracy sample preparation by Udo Linke.  Furthermore, we have
benefited  from helpful discussion with Joachim Krug. This work was
partially supported by the Fond der Chemischen Industrie, Germany.  TLE thanks the
Humboldt Foundation for generous support under a U.S. Senior Scientist Award and is also
grateful for funding from NSF-MRSEC, grant DMR 00-80008.

\vspace{.5cm}
\noindent * Corresponding author\\  Email: m.giesen@fz-juelich.de\\
Internet: http://www.fz-juelich.de/isg/isg3/Giesen/ag-giesen1.htm



\begin{thebibliography}{10}
\bibitem{noteE} Specifically, from use of a Lennard-Jones potential, Ref.\
\cite{E02} claims that the minimum stiffness occurs for steps in the $\langle 130
\rangle$-direction of Cu(001) and has simple sinusoidal variation proportional to
``crystalline anisotropy."
\end{thebibliography}
\end{document}